\documentclass[10pt, paper=a4, UKenglish]{article}
\usepackage{graphicx}
\def\Title#1{\begin{center} {\Large #1 } \end{center}}
\def\Author#1{\begin{center}{ \sc #1} \end{center}}
\def\Address#1{\begin{center}{ \it #1} \end{center}}

\newcommand\pubblock{\rightline{\begin{tabular}{l} Proceedings of the CTD/WIT 2019\\ \pubnumber\\
         \pubdate  \end{tabular}}}

\newenvironment{Abstract}{\begin{quotation} \begin{center} 
             \large ABSTRACT \end{center}\bigskip 
      \begin{center}\begin{large}}{\end{large}\end{center} \end{quotation}}

\newenvironment{Presented}{\begin{quotation} \begin{center} 
             PRESENTED AT\end{center}\bigskip 
      \begin{center}\begin{large}}{\end{large}\end{center} \end{quotation}}

\def\Acknowledgements{\bigskip  \bigskip \begin{center} \begin{large}
      \bf ACKNOWLEDGEMENTS \end{large}\end{center}}

\def\beq{\begin{equation}}
\def\eeq#1{\label{#1}\end{equation}}
\def\eeqn{\end{equation}}

\def\beqa{\begin{eqnarray}}
\def\eeqa#1{\label{#1}\end{eqnarray}}
\def\eeqan{\end{eqnarray}}

\let\bar=\overbar

\def\Dslash{\not{\hbox{\kern-4pt $D$}}}
\def\dslash{\not{\hbox{\kern-2pt $\del$}}}

\def\msb{{\bar{\ssstyle M \kern -1pt S}}}

\usepackage{color}
\usepackage{lineno}
\usepackage{subfig}
\usepackage{hyperref}

\newcommand\pubnumber{PROC-CTD19-006}

\newcommand\pubdate{\today}

\def\KAIST{
Department of Physics, \\
Korea Advanced Institute of Science and Technology, Republic of Korea}

\def\CAPP{
Center for Axion and Precision Physics Research, \\
Institute for Basic Science (IBS), Daejeon 34051, Republic of Korea}

\def\OSAKA{
Department of Physics, Graduate School of Science, \\
Osaka University, Toyonaka, Osaka 560-0043, Japan}

\newcommand{\conference}{Connecting the Dots and Workshop on Intelligent Trackers (CTD/WIT 2019)\\
Instituto de F\'isica Corpuscular (IFIC), Valencia, Spain\\ 
April 2-5, 2019}

\usepackage{fancyhdr}
\definecolor{mygrey}{RGB}{105,105,105}
\begin{document}


\large
\begin{titlepage}
\pubblock

\vfill
\Title{GPU Tracking in the COMET Phase-I Cylindrical Drift Chamber }
\vfill

\Author{Beomki Yeo} 
\Address{\KAIST}
\Author{Myeong Jae Lee, Yannis K. Semertzidis}
\Address{\CAPP}
\Author{Yoshitaka Kuno}
\Address{\OSAKA}
\vfill

\begin{Abstract}
The GPU-accelerated track finding method is investigated to track electrons from neutrinoless muon decay in the COMET Phase-I experiment. Inside the cylindrical drift chamber, one third of the signal electron trajectories are composed of multiple turns where the correct hit assignments to each turn partition are significant in the track finding. Scanning all possible track seeds of position and momentum can resolve the hit-to-turn assignment problem with a high robustness, but requires a huge computational cost: The initial track seeds $(\theta,z,p_x,p_y,p_z)$ have broad uncertainties, so there exists many number of seeds that should be compared. In this article, this problem of massive computations are mitigated with 1) the parallel computing of Runge-Kutta-Nyström  track propagation with the GPU, and 2) an initial guess on the seeds using the Hough transform and the detector geometry. The computation speed enhancement compared to the CPU is also presented.
\end{Abstract}

\vfill

\begin{Presented}
\conference
\end{Presented}
\vfill
\end{titlepage}
\def\thefootnote{\fnsymbol{footnote}}
\setcounter{footnote}{0}
%

\normalsize 


\section{Introduction}
\label{intro}

\begin{figure}[!h]
\centering
\begin{minipage}{.45\textwidth}
  \centering
  \includegraphics[width=1\linewidth]{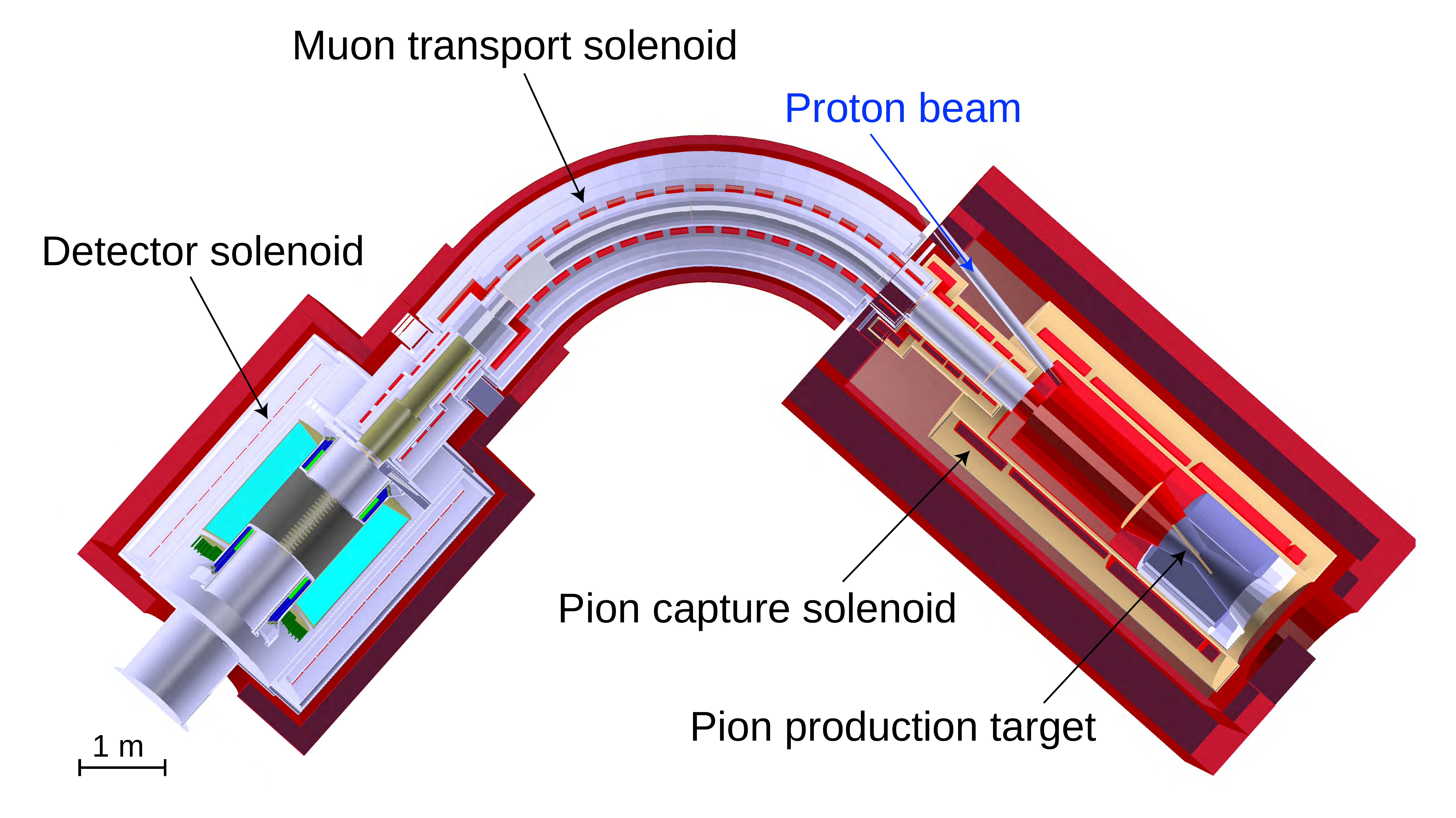}
  \vspace{5mm}
  \caption{COMET Phase-I layout.}
  \label{fig:COMETPhase1Layout}
\end{minipage}%
\quad
\begin{minipage}{.45\textwidth}
  \centering
  \includegraphics[width=1\linewidth]{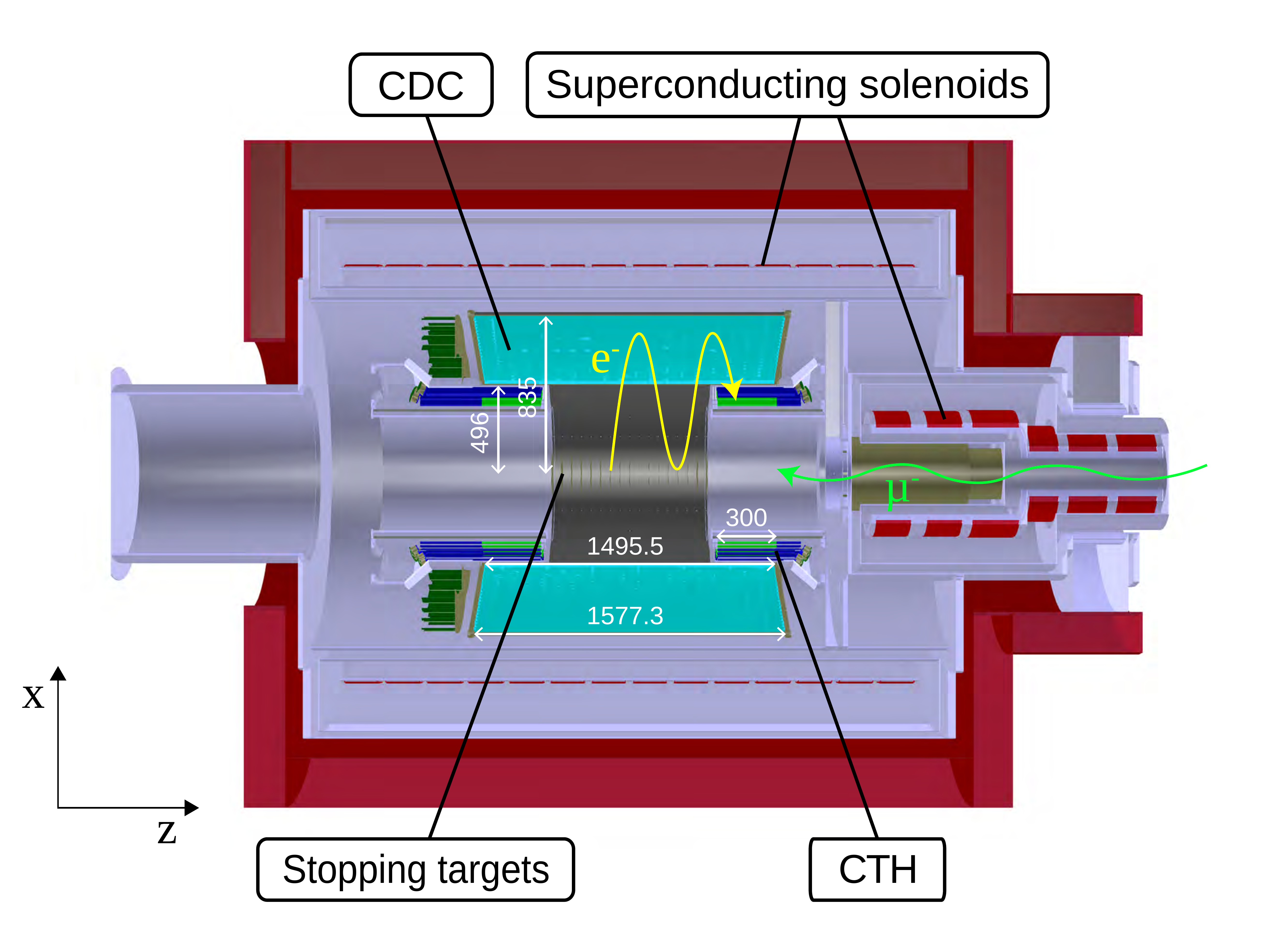}
  \caption{The layout of the CyDet. The yellow line represents the helix trajectory of an electron. The sizes are in mm unit.}
  \label{fig:CyDet}
\end{minipage}
\end{figure}

\indent The COMET experiment \cite{Adamov:2018vin}, located at J-PARC in Japan, will investigate the neutrinoless muon to electron conversion $(\mu^- + N \rightarrow e^- + N)$ in a muonic atom. Its single event sensitivities in Phase-I and Phase-II are $3.1 \times 10^{-15}$ and $2.6 \times 10^{-17}$, respectively, which are about 100 and 10000 times enhancement compared to the latest experimental upper limit of $7\times10^{-13}$ \cite{SINDRUM2}. In the Phase-I experiment (see Fig. \ref{fig:COMETPhase1Layout} for the layout), the proton beam from J-PARC accelerator hits the production target to generate the pions which decay into muons subsequently. Those muons are transported along the transport solenoid. Afterwards, they enter the detector solenoid region called CyDet and are stopped by the aluminum stopping targets to form muonic atoms. The stopped muons either decay into an electron and two neutrinos (Decay-In-Orbit, or DIO) or are captured by the nucleus. The neutrinoless muon to electron conversion is also possible with the aid of the neutrino oscillation while its branching ratio $(O(10^{-54}))$ is too low to observe. However, if new physics beyond the Standard Model exists, the neutrinoless muon to electron conversion may occur with an observable branching ratio. Meanwhile, the energy of the signal electrons from the muon conversion is fixed to 104.97 MeV in the case of the aluminum stopping targets, whereas electrons from DIO have a certain distribution whose endpoint energy is almost same with the signal energy \cite{Czarnecki}. Therefore the CyDet is required to have a good energy resolution to separate the signal electrons from the DIO electrons.

\begin{figure}[!htp]
  \centering
  \includegraphics[width=0.8\linewidth]{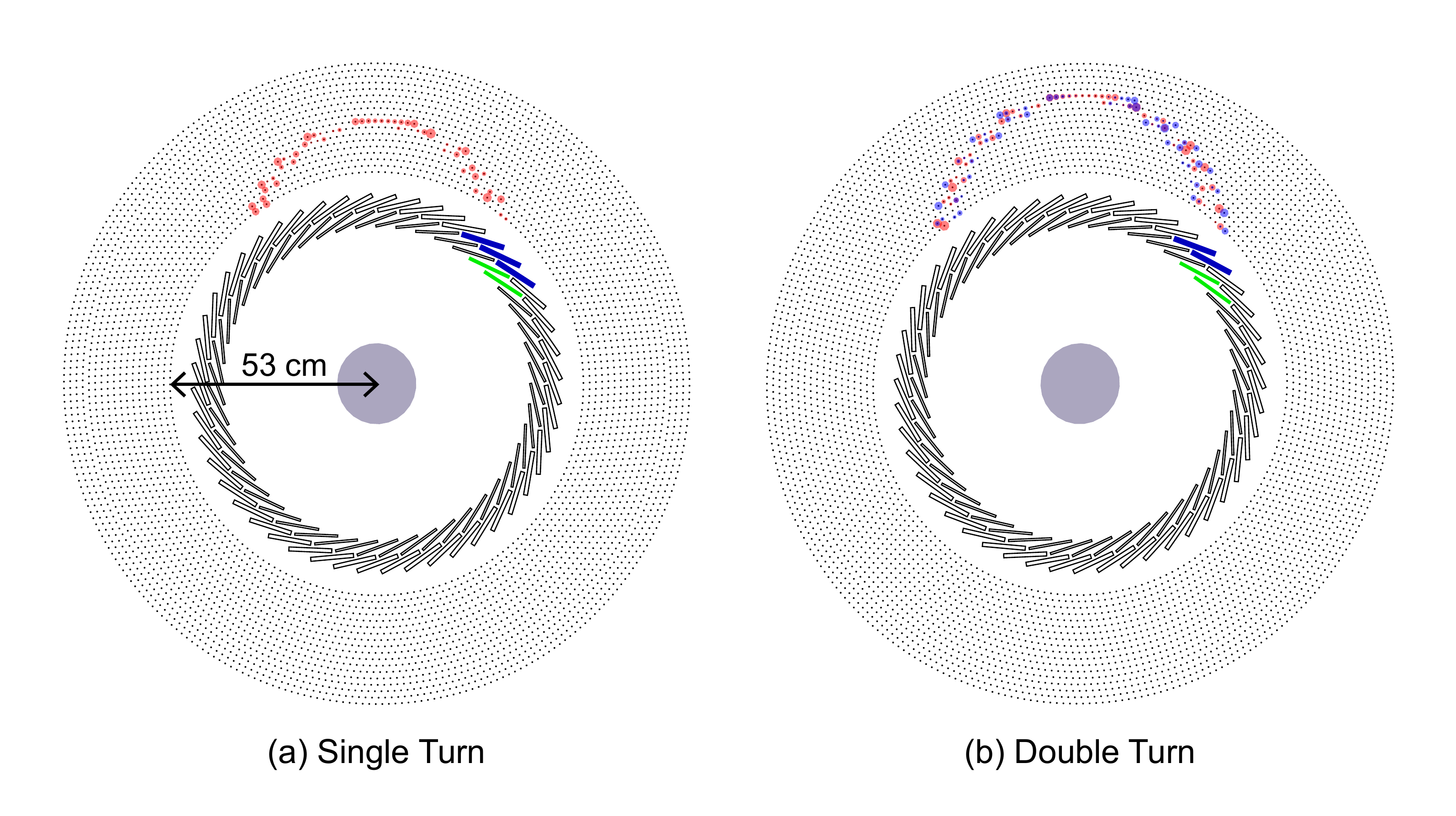}
  \caption{The XY projection of (a) the single turn event and (b) the double turn event. The transparent red and blue circles represent the hits from the first turn and second turn, respectively. The gray circle represent the muon stopping targets. The blue and green rectangles are triggered cherenkov and scintillators, respectively.}
  \label{fig:EventTypes}
\end{figure}

\indent The CyDet is composed of two detectors: the cylindrical drift chamber (CDC) and the CyDet trigger hodoscope (CTH). The electrons emitted from the stopping targets make a helix under the 1T magnetic field and ionize the gas molecules while passing through the CDC filled with the gas mixture of He (90 \%) and $\textrm{C}_4\textrm{H}_{10}$ (10 \%). The induced currents from the ionized electrons and ions are measured by the sense wires arranged across the eighteen layers with stereo configuration. The CTH is located at the ends of the CDC and consists of the scintillators and cherenkov detectors to distinguish the electron events. Figure \ref{fig:CyDet} shows the layout of the CyDet with an electron that propagates in a helix trajectory and reaches the CTH. \\
\indent There are two event types for electrons in terms of the number of helix turns: a single turn and multiple turn event as displayed in Fig. \ref{fig:EventTypes} with the XY projection. The purpose of event reconstruction in the COMET experiment is finding the momentum of the first turn partition of the electron trajectory where the energy loss is minimal. However, the track finding of the multiple turn events require the hit classification for each turn, and this requires huge computation because there exists many hit combinations to be checked.

\indent In this proceeding, we introduce the track finding method which scans the set of the possible tracking seeds, rather than the set of hit  combinations, and generates the Runge-Kutta-Nyström (RKN) tracks from the seeds to classify the hits nearby. Since the number of seeds to be scanned is still not small, the algorithm was parallelized  using the CUDA language \cite{CUDA} to achieve the reasonable computing speed with the GPU. The article is organized as follows: In Section \ref{sec:GPGPUScanningMethod}, the seed scanning method will be explained for the algorithm and its CUDA implementation. Section \ref{sec:Results} will show the results of the method. Section \ref{sec:TechonicalAspects} will introduce some technical aspects regarding the GPU implementation. Conclusions and outlooks will follow at Section \ref{sec:Conclusions}.

\section{Parallelized Seed Scanning with Runge-Kutta-Nyström Track}
\label{sec:GPGPUScanningMethod}

\subsection{Preparing the set of the seeds}
\begin{figure}[!h]
	\centering
	\begin{minipage}{.45\textwidth}
  	\includegraphics[width=1\linewidth]{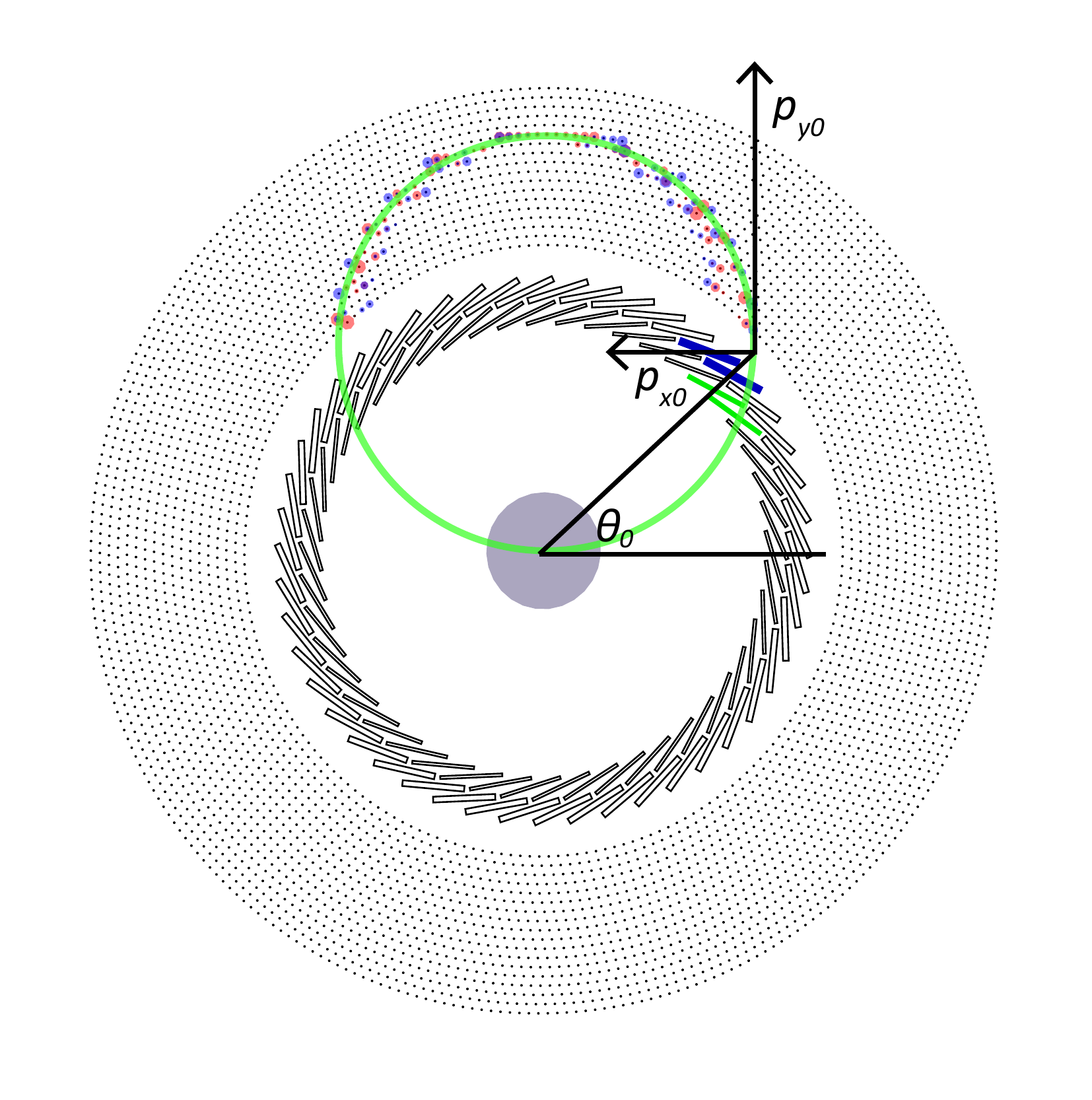}
  	\caption{Extracting transverse seeds from the (green) circle found by the Hough transform.}
  	\label{fig:CircleFitting}  
  	\end{minipage}
  	\quad
  	\begin{minipage}{.45\textwidth}
  \includegraphics[width=1\linewidth]{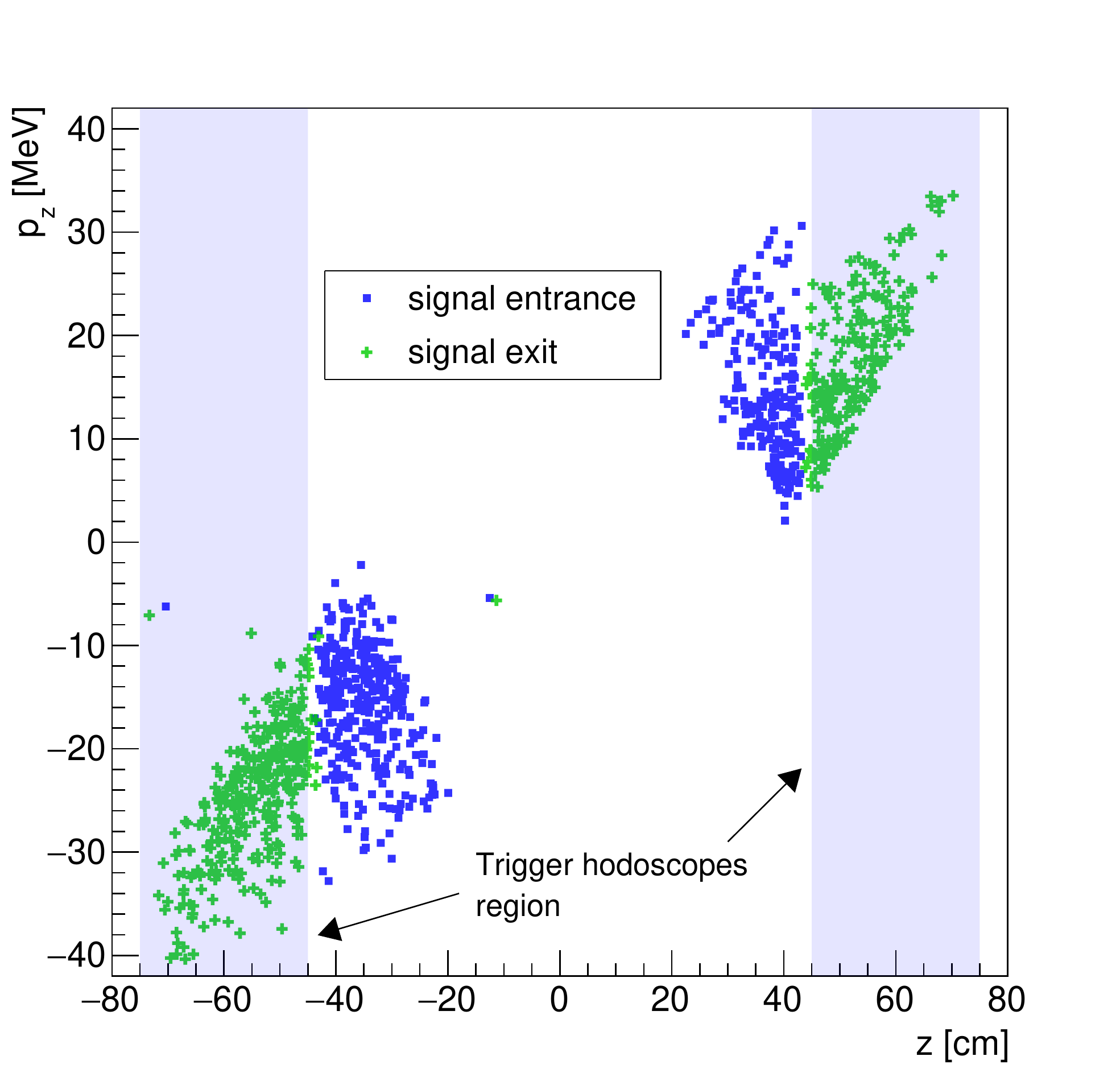}
  \caption{The $(z_0,p_{z0})$ distribution of the signal electrons}
  \label{fig:zpz_pattern_multi}
  	\end{minipage}  
\end{figure}

The seed of the track is supposed to have six track parameters: three for the position $(R_0, \theta_0, z_0)$ in the cylindrical coordinate and three for the momentum $(p_{x0}, p_{y0}, p_{z0})$ in the cartesian coordinate. Since we define the seed in front of the first layer of CDC where the radius component $(R_0)$ is fixed to 50 cm, the actual number of parameters to be scanned is five. The scanning range of the transverse seeds $(\theta_0, p_{x0}, p_{y0})$ was obtained by the circle finding method, namely the Hough transform (see Fig. \ref{fig:CircleFitting}). The resolution of the Hough transform was about $0.01$ radian for $\theta_0$ and $2.5$ MeV for $p_{x0}$ and $p_{y0}$, respectively. For the longitudinal seeds, we utilized the fact that triggering tracks always end at the CTH, which means that the last turn partition should have a certain pattern in $(z_0, p_{z0})$. Figure. \ref{fig:zpz_pattern_multi} shows the distribution of $(z_0, p_{z0})$ for the last turn partition both for the CDC entrance and exit. Since all components in the seeds have the known ranges for the last turn partition, the track finding starts from there. The total scanning range was set to cover at least 90\% of the monte-carlo truth values of the CTH triggering event samples, and the granularity of the seeds was set not to exceed the GPU memory capacity. With these conditions, the number of seeds was about few tens of thousand.

\subsection{The seed scanning method and its implementation on the GPU}
\label{subsec:ScanningMethod}
\begin{figure}[!h]
    \centering
    \includegraphics[width=.3872\linewidth]{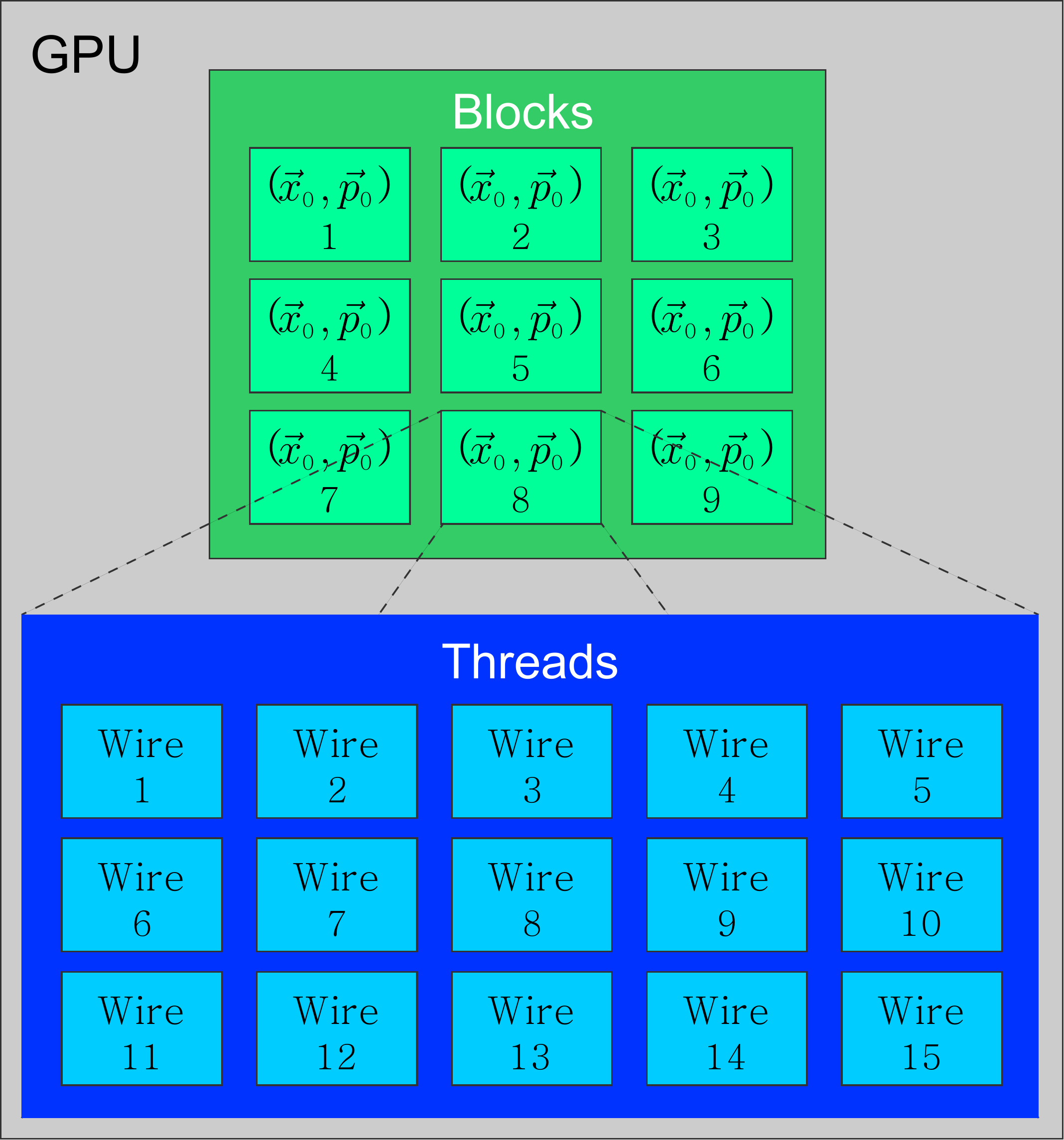}
  	\quad
    \includegraphics[width=.57\linewidth]{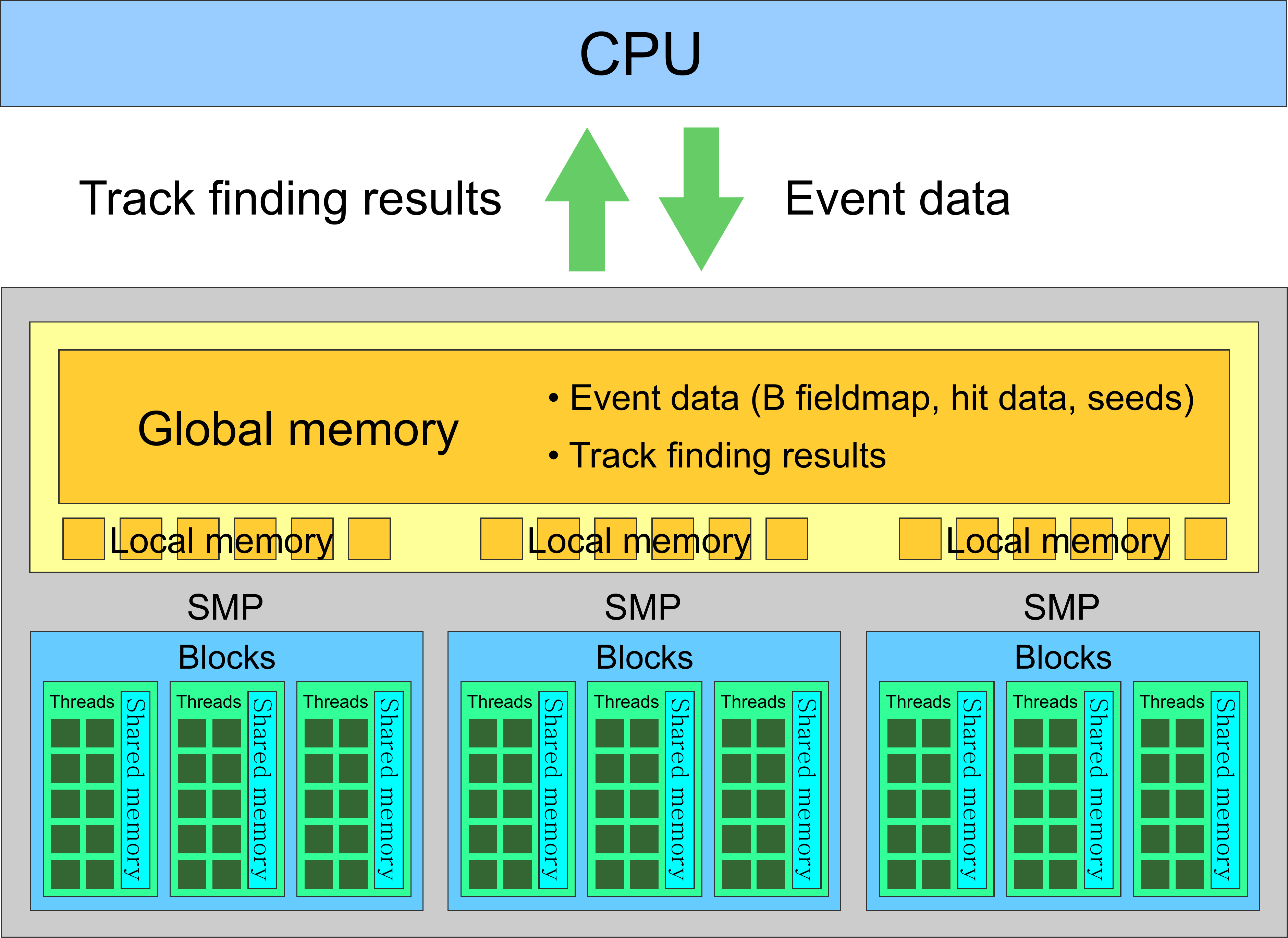}
    \caption{(left) The schematic view of the seed scanning method with the GPU. $(\vec{x}_0, \vec{p}_0)$ represents the seed. (right) The schematic view of the memory hierarchy. SMP stands for the streaming multiprocessor.}
  \label{fig:GPUstructure}
\end{figure}

The seed scanning method is a process that iterates over the seed set to select the best seed: For every iteration, a track is extrapolated from the seed to find the distance of the closest approach (DCA) between the extrapolated track and hits. The RKN method was used for the track parameter propagation \cite{STEP}. Then, the goodness of each seed was evaluated based on the chi-square like value $(E)$ defined by:
\begin{equation}
    E = \sum_{\textrm{wires}}\textrm{min}(d_L^2, d_R^2, \lambda),
\end{equation}
where $d_L$ and $d_R$ are DCA to the left and right hit of the wire, respectively, which are from the left-right ambiguity of the wire measurement. $\lambda$ is the cutoff value to ignore the contribution of hits from the other turn partitions. After calculating $E$ for all seeds, we took the several candidate seeds with the lowest $E$, and make other seed sets around them with finer granularity. Afterwards, the same scanning process was repeated ten times to get enough resolution for the seed selection. The hits whose square of DCA is less than the cutoff value $(d^2<\lambda)$ were classified as the hits of the turn partition being investigated. \\ \indent
Because the effect from the multiple scattering gets more significant as the extrapolation length gets longer, the track finding with a  bidirectional extrapolation was adopted to mitigate the material effect. It means that the scanning method was done independently for the seed sets collected at both the CDC entrance and exit. The quality cut that the number of hits commonly found from the both direction $(N_c)$  should be larger than or equal to 20 was applied to remove the misleading track finding results. Afterwards, the tracking finding results including the best seed and the hit classification were used for the track fitting with the Kalman filtering method \cite{GENFIT2}. \\
\indent The computation of the GPU is performed by the threads, which are the smallest software units in parallel, through the CUDA kernel function. The threads are grouped into a block, and the streaming multiprocessors of the GPU can operate the multiple units of blocks concurrently. In the seed scanning method, each tracking seed is delivered to each block, and each thread in a block calculates the DCA for each wire (see Fig. \ref{fig:GPUstructure}) Therefore, the number of blocks to be launched in the GPU is same with the number of scanning seeds, and the number of threads in a block is same with the number of the fired wires. Before launching the kernel function, the event information such as the magnetic field, the hit data, and the seeds is transferred to the global memory of the GPU, which has a few of GB capacity and can be accessed from any thread in any block. The track finding results are also stored at the global memory, and the GPU transfer them back to the CPU. A disadvantage of the global memory is that it has a small memory access speed compared to the CPU. The other variables generated during the RKN method are stored at local memory whose access is confined to a specific thread but still as slow as global memory. Even though there is a shared memory residing on the GPU chip whose access speed is comparable to the CPU, it was not exploited as more shared memory allocation leads less number of blocks being launched simultaneously. The K40m and K80 model of the NVIDIA Tesla family \cite{TESLA} were tested to benchmark the performance.

\subsection{Full event reconstruction}
\begin{figure}[!h]
	\centering
	\includegraphics[width=1.\linewidth]{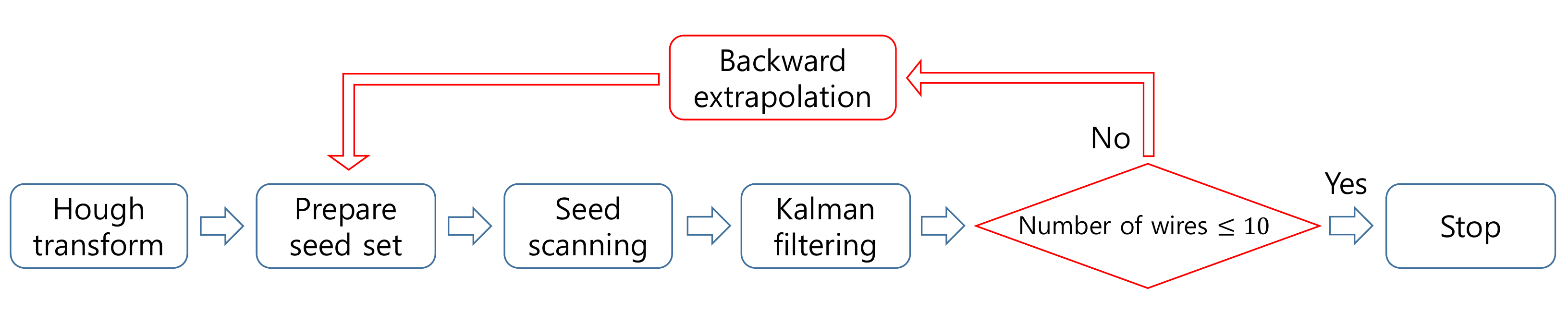}
	\caption{Flowchart for the event reconstruction process}
	\label{fig:Event_reconstruction_process}
\end{figure}

For the neighboring previous turn partition, the track finding and fitting process are same with that of the last turn partition except the preparation of the longitudinal seed $(z_0,p_{z0})$. The longitudinal seeds at the entrance and the exit were predicted by backwardly extrapolating the track that was already fitted. The resolution of $z_0$ and $p_{z0}$, were 2.6 cm and 1.7 MeV, respectively. The whole event reconstruction process stops when the number of remaining unclassified hits becomes less than ten. The flowchart for the event reconstruction process is illustrated at Fig. \ref{fig:Event_reconstruction_process}.

\section{Results}
\label{sec:Results}
\subsection{Track finding results}
\begin{figure}[!h]
\centering
\begin{minipage}{.45\textwidth}
  \centering
  \includegraphics[width=1.\linewidth]{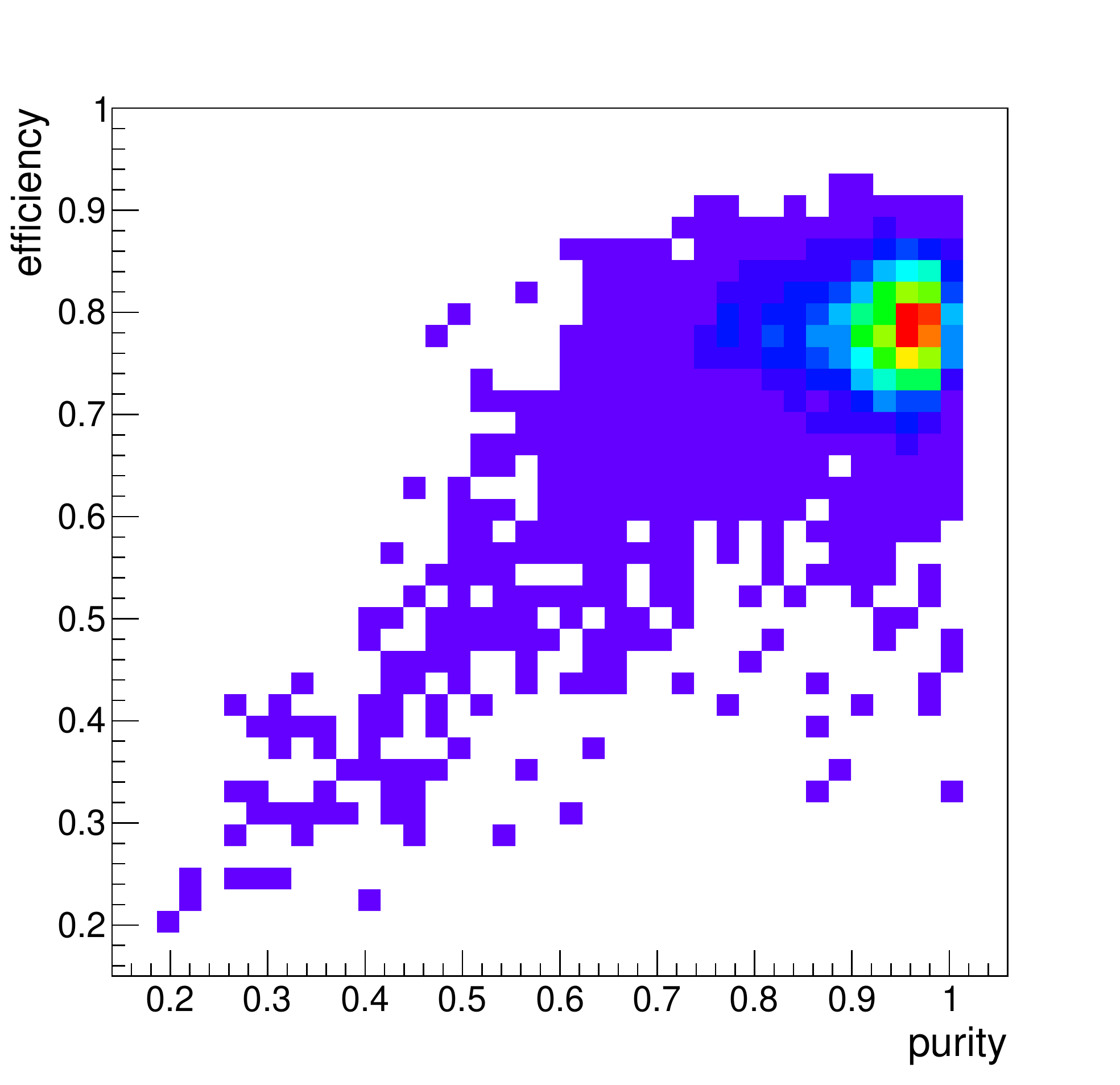}
  \caption{The two dimensional histogram of the efficiency and purity of the last turn partition.}
  \label{fig:EffPur}
\end{minipage}%
\quad
\begin{minipage}{.45\textwidth}
  \centering
  \includegraphics[width=1.\linewidth]{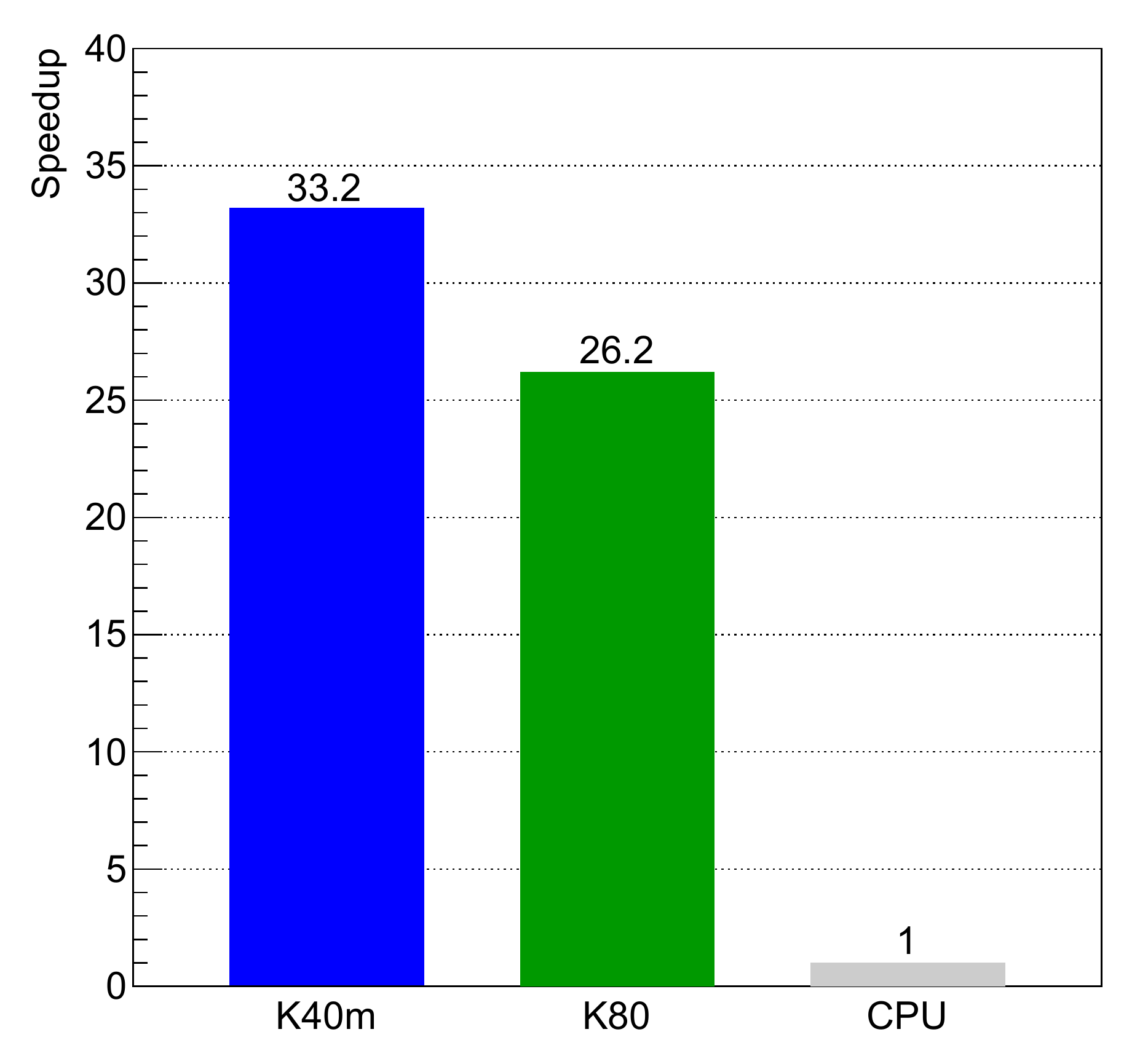}
  \caption{The speedups of the GPUs (K40m and K80) beyond the CPU (E5-2630).}
  \label{fig:CPUvsGPU}
\end{minipage}
\end{figure}

The quality of the track finding with the scanning method can be quantified by the classification efficiency $(\epsilon)$ and purity $(\pi)$ for the $n'$-th turn partition:
\begin{eqnarray}
    \epsilon_{n'} & = & \frac{C_{n'\rightarrow n'}}{C_{n'}}, \\
    \pi_{n'}      & = & \frac{C_{n'\rightarrow n'}}{\sum_{i=1}^{n}C_{i\rightarrow n'}},
\end{eqnarray}
where $n$ is the number of turns for an event, $C_{n'}$ is the total number of the hits from the $n'$-th turn partition, and $C_{i\rightarrow n'}$ is the number of the hits of the $i$-th turn partition classified as the hits of the $n'$-th turn partition. For example, the two dimensional histogram of Fig. \ref{fig:EffPur} shows $\epsilon_n$ and $\pi_n$ of the last turn partition after applying the quality cut of $N_c \geq 20$. The averages of $\epsilon_n$ and $\pi_n$ were 76 \% and 90 \%, respectively. \\
\indent The speed of scanning method in the GPU was compared with the CPU by serializing the same algorithms. As shown in Fig. \ref{fig:CPUvsGPU}, it turned out that the speed of K40m and K80 was 33 and 26 times faster than the CPU (Intel E5-2630), respectively. 

\subsection{Track fitting results}
\begin{figure}[!h]
  \centering
  \includegraphics[width=.49\linewidth]{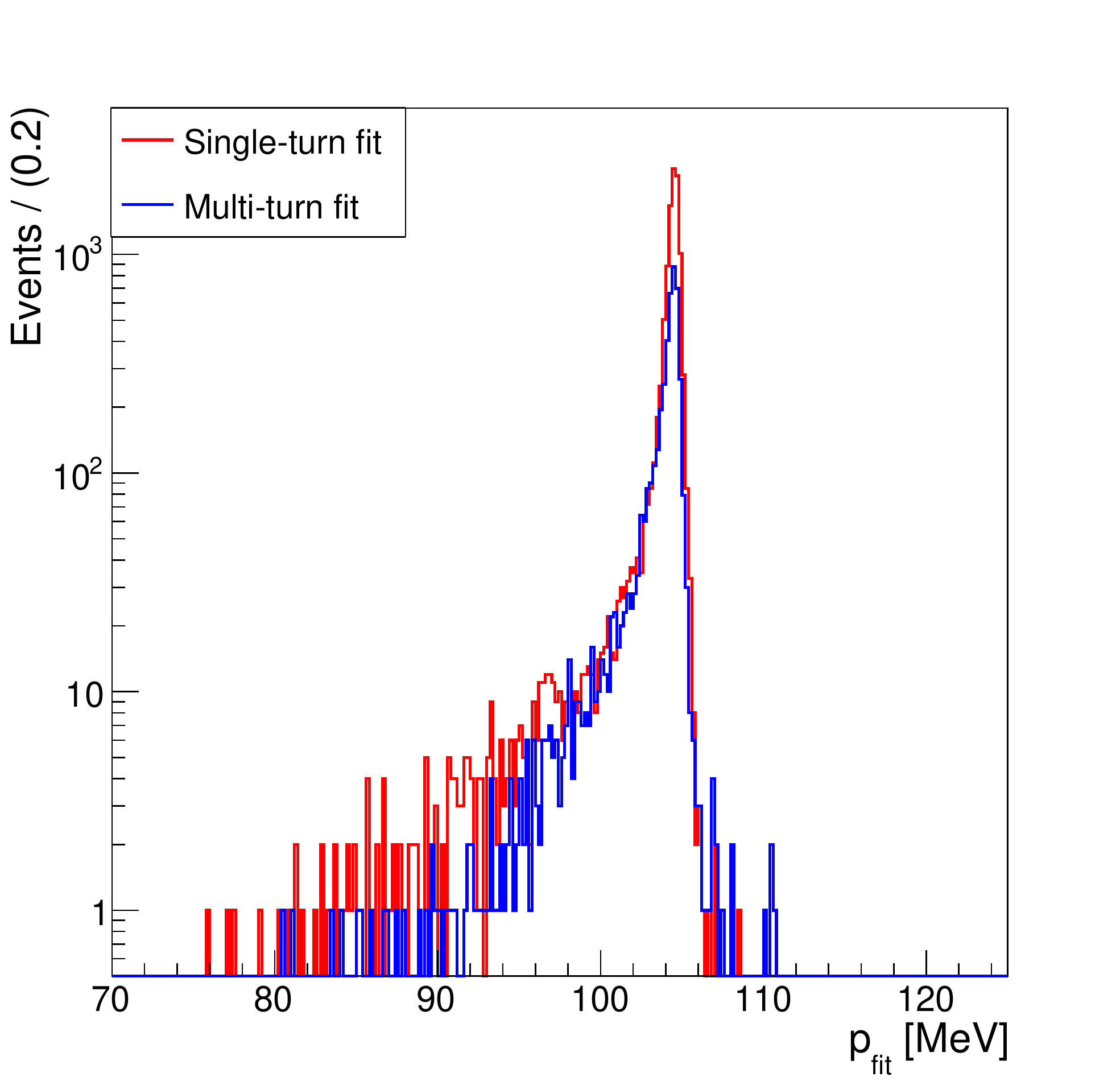} 
  \includegraphics[width=.49\linewidth]{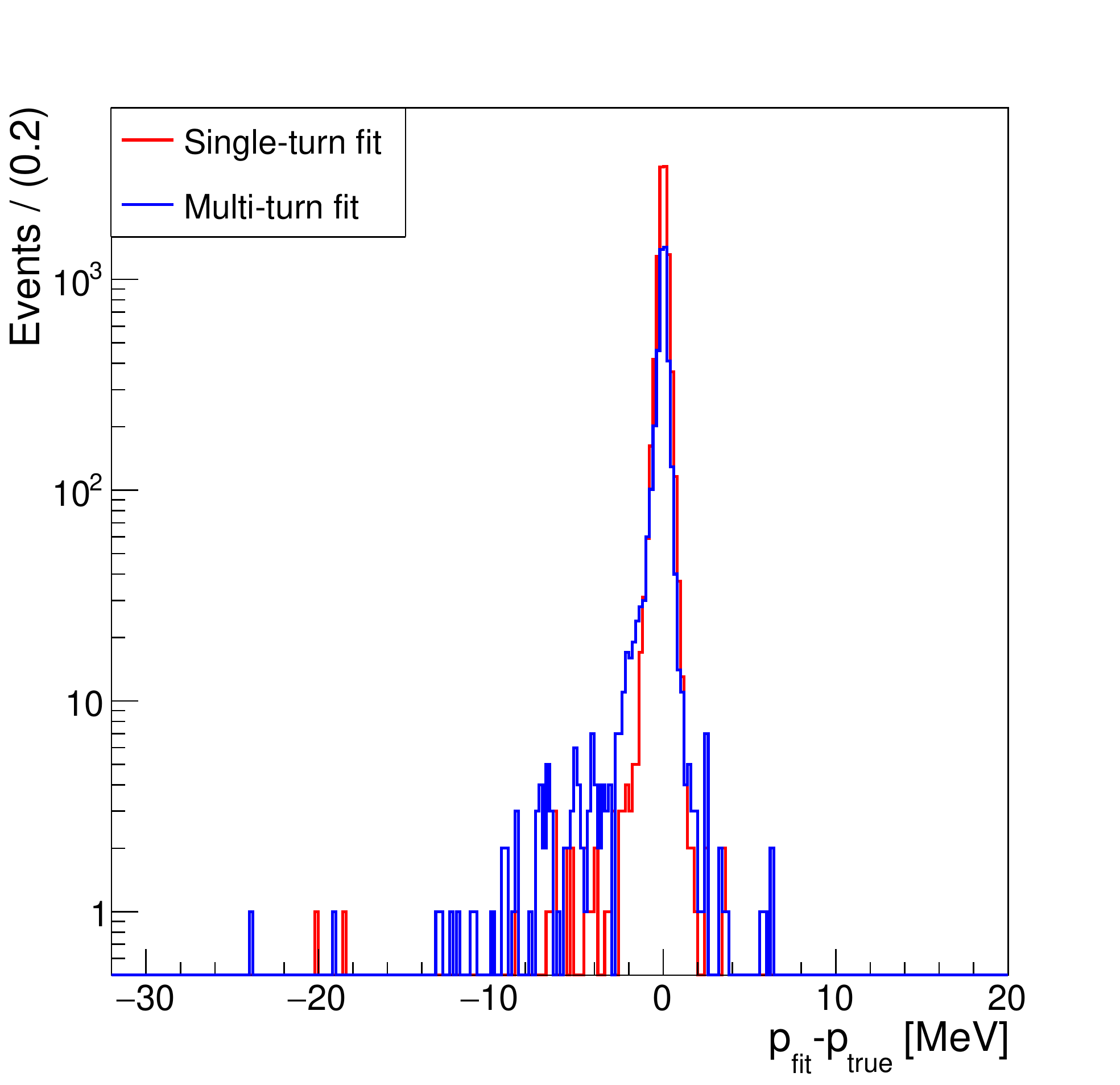}
    \caption{(left) The histogram of the fitted momentum and (right) its residual against the truth momentum at the first CDC entrance. The red and blue histogram represent the single turn events and multiple turn events, respectively.}
  \label{fig:fittedP}
\end{figure}

The track fitting results shown at Fig. \ref{fig:fittedP} represent the momentum distribution of the first turn partition, or the closest one to the first turn partition. Four of track quality cuts were given: 1) CTH triggering, 2) $N_c \geq 20$, 3) $\textrm{NDF} \geq 35$, and 4) $\chi^2/\textrm{NDF} \leq 2$. In the momentum residual histogram, the energy resolution for the core part was about 300 keV with the gaussian fitting. The tracking efficiency was defined as the number of events that pass the quality cuts divided by the number of CTH triggering events whose averaged number of hits per turn is larger or equal to 35. With this definition, the tracking efficiency for the single turn events and multiple turn events were 83\% and 75\%, respectively. Meanwhile, a tail in low energy region can be found in the histogram of the multiple turn events, compared to the single turn events. It is because some events were failed in reconstructing the first turn event leading to more energy loss in the finally fitted turn partition.

\section{Technical Aspects in the GPU-implementation}
\label{sec:TechonicalAspects}
\subsection{Limiting factors}
\begin{figure}[!h]
  \centering
	\includegraphics[width=.58\linewidth]{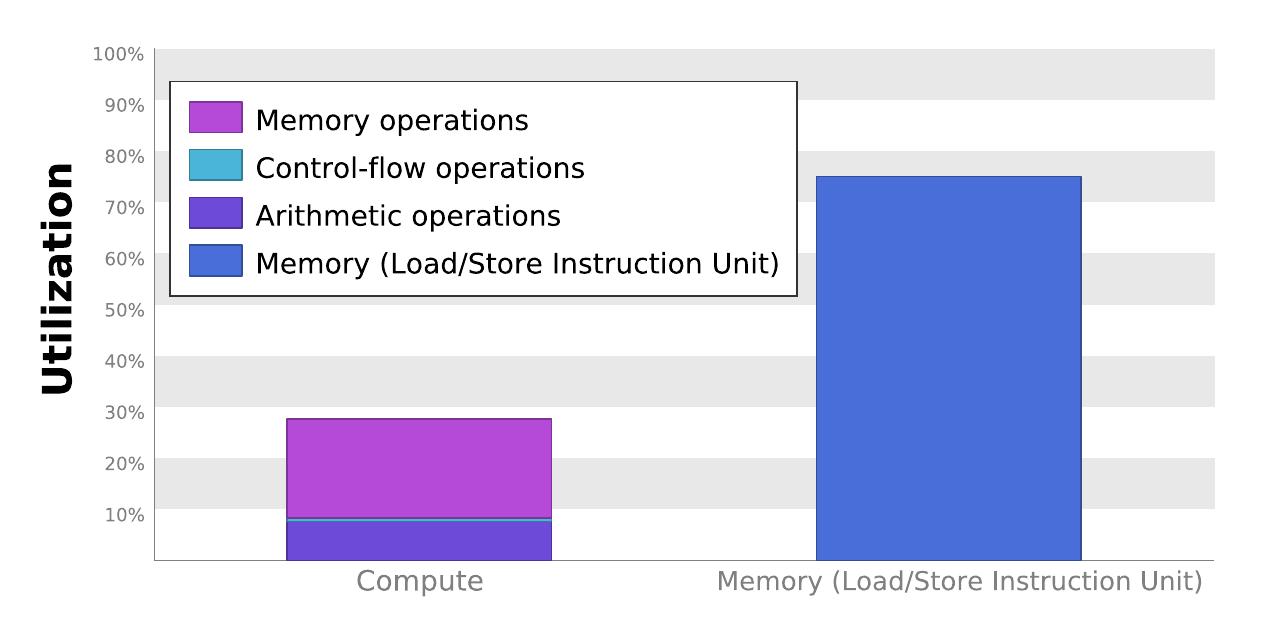} 
	\includegraphics[width=.40\linewidth]{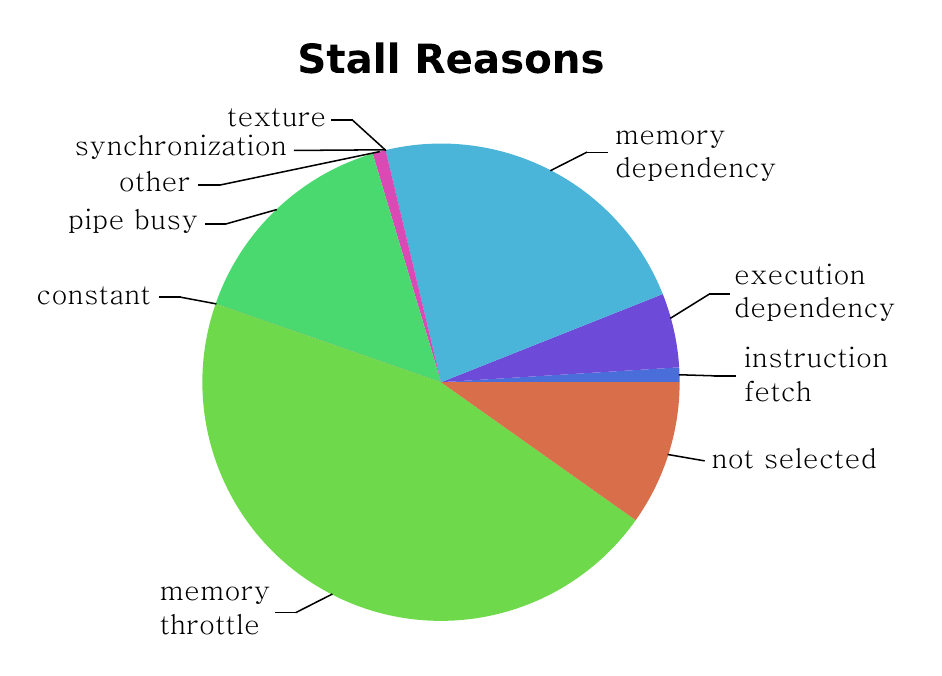}
	\caption{(left) The GPU utilization status for each category and (right) the stall reasons in the GPU, generated by NVIDIA Visual Profiler \cite{NVPROF}.}
	\label{fig:GPUutilization}	
\end{figure}

The performance of the GPU usually gets better when more GPU resources are occupied, and the occupancy is proportional to the number of the fired wires of the events, corresponding to the number of threads. Therefore, the occupancy is low for the events with small number of the hits. Another issue is the branch divergence where some threads do not operate while the others do, which lowers the efficiency of the GPU utilization. The branch divergence occurs when threads do not have the same branching behavior, and it is inevitable due to the feature of the RKN method where some threads can finish their extrapolations earlier, or later than others. The other issue, the main limiting factor, is the memory throttle occurred by the large memory bandwidth during the scanning method. Figure \ref{fig:GPUutilization} shows the GPU utilization status for each category and stall reasons, indicating that the bottleneck in the memory transaction is slowing the whole process.

\subsection{Kernel function separation}
\label{subsec:GPUutilization}
To reduce the branching divergence, the DCA finding algorithm with the RKN method was separated into two kernel functions. In the first kernel function, each seed was delivered to each thread, and recorded the track parameters of all steps during the RKN extrapolation to the end of the turn partition. The algorithm of the second kernel function  which finds DCA to the wires is same with the description in Section \ref{subsec:ScanningMethod}, except that the track propagation starts from the recorded track parameter which is closest to the corresponding wire. \\

\section{Conclusions and Outlooks}
\label{sec:Conclusions}
The GPU-accelerated track finding algorithm for the COMET Phase-I experiment was introduced for the event reconstruction of the multiple turn electrons. The seed scanning method followed by the Kalman filtering showed the acceptable energy resolution of 300 keV. However, the further optimization of many parameters such as the cutoff value ($\lambda$) or the granularity of the seed sets is required to increase tracking efficiency and resolution. The background events of DIO, radiative muon capture and radiative pion capture as well as the other new physics processes such as $\mu^- \rightarrow e^+$ \cite{Beomki} will also be investigated.


\Acknowledgements
This work was supported by IBS-R017-D1 of the Institute for Basic Science, the Republic of Korea (B.Y., M.L., Y.S.) and the JSPS KAKENHI Grant No. 18H04231 (Y.K.)


\end{document}